\documentclass[twocolumn]{article}

\usepackage[utf8]{inputenc}
\usepackage{graphicx}
\usepackage{amsmath,amssymb}
\usepackage{dblfloatfix} 
\usepackage{hyperref}
\usepackage{orcidlink}
\usepackage{enumitem}
\newenvironment{compactitemize}
  {\begin{itemize}[itemsep=0pt, topsep=1pt, parsep=0pt, partopsep=0pt]}
  {\end{itemize}}
\setlist[itemize]{
  itemsep=0pt,        
  parsep=0pt,         
  topsep=3pt,         
  partopsep=0pt
}
\usepackage[T1]{fontenc}
\usepackage{libertine}      
\usepackage[libertine]{newtxmath} 
\usepackage[backend=biber,style=numeric,sorting=nyt]{biblatex}
\renewbibmacro*{urldate}{}
\renewbibmacro*{url+urldate}{%
  \printfield{url}%
}
\usepackage{graphicx}
\usepackage{titlesec}
\titlespacing*{\section}
  {0pt}    
  {2ex}    
  {1ex}    

\titlespacing*{\subsection}
  {0pt}    
  {1.5ex}  
  {0.5ex}  

\usepackage{array} %
\usepackage{caption} 
\newcolumntype{P}[1]{>{\raggedright\arraybackslash}m{#1}}
\usepackage{cuted}
\usepackage[most]{tcolorbox}
\tcbset{
  frame code={},
  center title,
  boxsep=4pt,
  left=6pt,
  right=6pt,
  top=6pt,
  bottom=6pt,
  colback=gray!5,
  colframe=gray!40,
  coltitle=black,
  fonttitle=\bfseries,
}

\title{Real-World AI Evaluation: How FRAME Generates Systematic Evidence to Resolve the Decision-Maker's Dilemma}

\author{%
  \begin{tabular}{c@{\hspace{10em}}c}
    Reva Schwartz\,\orcidlink{0000-0002-9012-6306} &
    Gabriella Waters\,\orcidlink{0009-0001-1821-6091} \\
\href{mailto:reva@civitaas.com}{reva@civitaas.com} &
    \href{mailto:gwaters@vsu.edu}{gwaters@vsu.edu} \\
    Civitaas Insights &
    Center for Responsible AI, \\
    &
    Virginia State University, \\
    &
    and Civitaas Insights \\
  \end{tabular}
  \date{}
}

\begin{document}

\twocolumn[
\begin{@twocolumnfalse}
\maketitle
\begin{abstract}
Organizational leaders are being asked to make high‑stakes decisions about AI deployment without dependable evidence of what these systems actually do in the environments they oversee. The predominant AI evaluation ecosystem yields scalable but abstract metrics that reflect the priorities of model development. By smoothing over the heterogeneity of real‑world use, these model-centric approaches obscure how behavior varies across users, workflows, and settings, and rarely show where risk and value accumulate in practice. More user‑centric studies reveal rich contextual detail, yet are fragmented, small-scale and loosely coupled to the mechanisms that shape model behavior. The Forum for Real‑World AI Measurement and Evaluation (FRAME) aims to address this gap by combining large‑scale trials of AI systems with structured observation of how they are used in context, the outcomes they generate, and how those outcomes arise. By tracing the path from an AI system’s output through its practical use and downstream effects, FRAME turns the heterogeneity of AI‑in‑use into a measurable signal rather than a trade‑off for achieving scale. The Forum establishes two core assets to achieve this: a Testing Sandbox that captures AI‑in‑use under real workflows at scale and a Metrics Hub that translates those traces into actionable indicators.

\end{abstract}
\end{@twocolumnfalse}
\vspace{2em} 
]

\section{FRAME Purpose and Activities}
\subsection{Background}
Recent reporting shows that broad‑scale generative AI deployments have produced uneven results, with many initiatives stalling or failing to deliver meaningful value \cite{challapally_genai_divide_2025}. Policymakers and organizational leaders need concrete expectations about what these deployments deliver in their own settings so they can make comparisons across sites and sectors\cite{mcalister_rethinking_2025,el_arab_bridging_2025,subasri_detecting_2025}. Today, these stakeholders have only a small set of metrics to rely on, typically leaning on benchmarks of AI model capabilities because they are scalable and quantitative. Yet even as AI systems are being globally adopted and embedded in everyday life, benchmarks remain anchored in testing paradigms that bear little resemblance to real‑world use. The benchmarking ecosystem is built to answer questions about optimization and safety in model development, with performance metrics based on standardized tasks under optimized conditions. Since it produces scant evidence about how AI tools are actually taken up, worked around, or ignored in practice, existing benchmarks map poorly onto deployment‑level questions about adoption, safeguards, oversight, and distributional impacts. This contributes to a structural mismatch between the evidence the ecosystem produces and the evidence that social, economic, and cultural decision‑makers need.

A key weakness of benchmarking is its inability to model or account for the unpredictable and variable ways people leverage AI technology in context, including how they express needs, interpret and act on outputs, and adapt and repurpose systems. Gabriella Waters has coined this variability as “user entropy,” and it can be a significant factor in whether a deployment succeeds or fails\cite{jahani_as_2024,atil_non-determinism_2025,pine_politics_2015,salaudeen_measurement_2025,wallach_position_2025}. \textit{Model‑centric} evaluations essentially treat user entropy as statistical noise to be eliminated rather than as a primary measurement signal\cite{bean_measuring_2025,nair_critical_2025,wang_critical_2026}. \textit{User‑centric} studies generate complementary, context‑rich evidence, but usually within narrow sites or populations, and disconnected from dominant benchmarking pipelines\cite{berman_scoping_2024,Ibrahim2025InteractiveEvaluations,matias_humans_2023}. These two strands of evaluation—scalable model‑centric tests and local user‑centric pilots—are rarely connected systematically, leaving decision‑makers without evidence that both reflects the diversity and unpredictability of real use and remains comparable across many deployments.

Current limitations of connecting model behavior to the real-world uses, failures, and adaptations that drive AI’s higher‑order effects\cite{schwartz_reality_2025} place leaders in a “decision‑maker’s dilemma”: they must reinterpret abstract scores for the contexts they oversee, without clear evidence about where risk and value lie or how they are likely to manifest\cite{capraro_impact_2024,maeda_when_2024,valenzuela_how_2024,tang_between_2025}. Just as clinical trial results cannot tell a city planner how a virus will move through a particular school system, benchmark scores and lab tests do not show chief information officers how large language models will alter workflows in their own organizations\cite{capraro_impact_2024,maeda_when_2024,valenzuela_how_2024,tang_between_2025}. Dashboards and trainings can help people interpret such metrics, but they cannot substitute for systematic evidence about how AI shapes everyday work, culture and the broader society. Figure~\ref{fig:user-entropy-dilemma} shows how modeling user entropy can help to connect abstract metrics to concrete deployment decisions.

\begin{strip}
  \centering
  \includegraphics[width=0.8\textwidth]{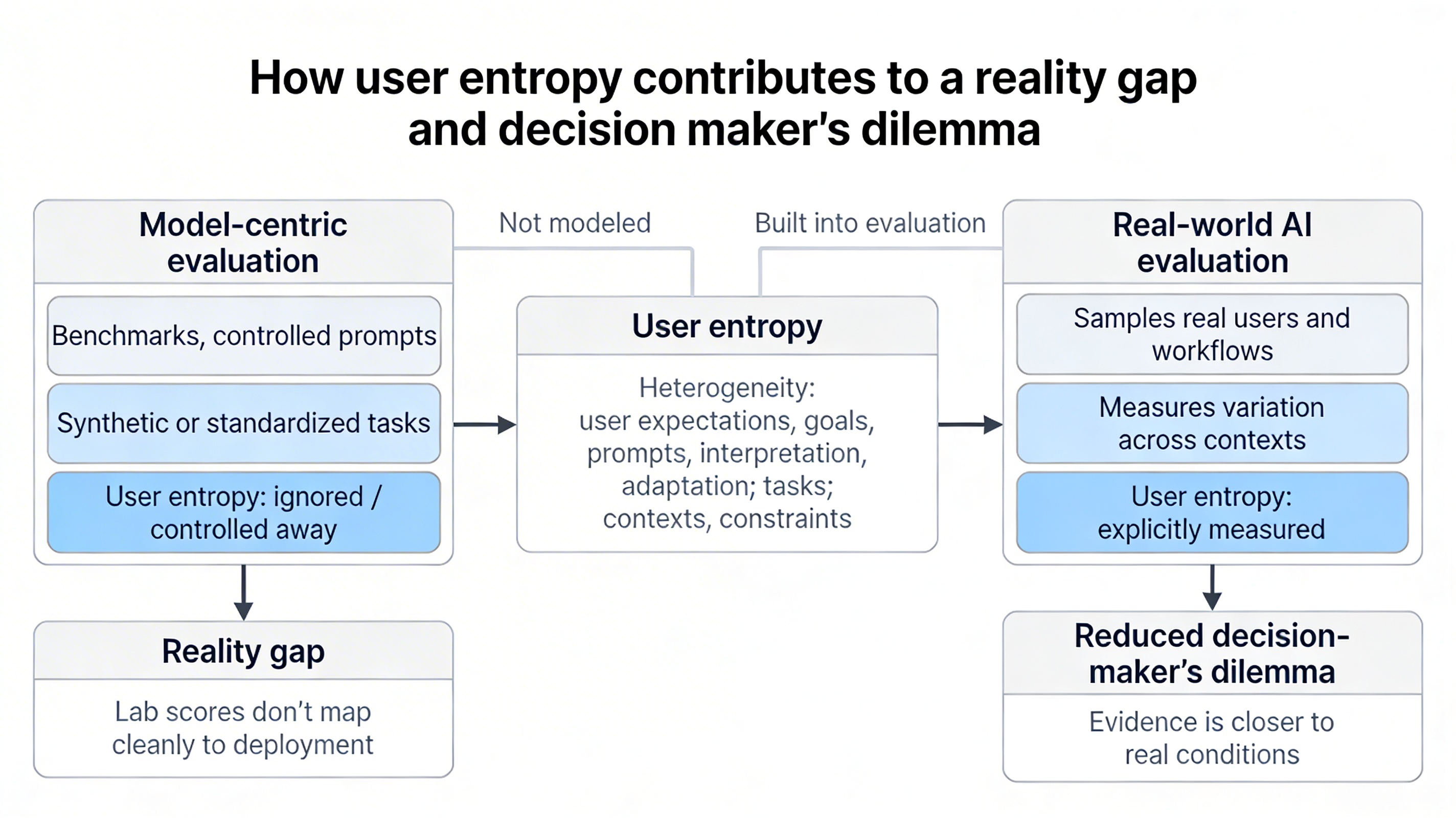}
  \captionof{figure}{Modeling user entropy can support concrete AI deployment decisions. (Generative artificial intelligence was used to support the creation of this graphic representing the authors’ own ideas, data, and words on this topic.)}
  \label{fig:user-entropy-dilemma}
\end{strip}

\subsection{The Next Generation of AI Evaluation}
Anchored at Virginia State University, the Forum for Real‑World AI Measurement and Evaluation (FRAME) was established to address these challenges by measuring system behavior in real contexts, not just on optimized tests. FRAME's membership forms a global, interdisciplinary coalition spanning measurement science, machine learning, social science, and the humanities across academia, industry, and government. Instead of each organization building its own costly testing stack, the Forum runs large‑scale evaluations--supported by sponsoring organizations--so sectors can assess AI in conditions similar to their own without exposing proprietary data. FRAME also acts as a "methods lab" by pioneering approaches to capture user entropy at scale and translate observations into comparable indicators across sites. 

FRAME evaluations complement existing capability benchmarks, reinforcement learning from human feedback (RLHF) pipelines, and adversarial testing. While those tools illuminate model behavior and task performance, FRAME foregrounds user entropy to clarify how AI actually plays out in real‑world use. This gives decision‑makers evidence to interpret what AI‑in‑use means for their own purposes and to align deployments more closely with institutional goals and public benefit. To conduct this work, FRAME is developing a centralized research and evaluation infrastructure built around two core components:

\begin{itemize} 
\item The \textbf{testing sandbox} is a controlled but realistic environment that uses large-scale remote participant panels to evaluate commercial AI systems under task-driven scenarios. Rather than relying on people as labelers, panelists provide descriptive accounts of how they leverage, repurpose, or abandon AI tools, offering a clear window into user entropy in practice. The sandbox maintains strict human‑subjects protections and relies on carefully designed proxy tasks to safely measure high-stakes risks without exposing participants to harm or sensitive content. 
\item The \textbf{metrics hub} is a translation layer that converts sandbox evaluation outcomes into decision-ready evidence about AI system behavior with real users in real contexts. It produces key indicators of system utility, friction, and resilience—offering a grounded alternative to abstract leaderboard scores. Released on a regular cadence, these indicators align with stakeholders’ deployment needs and complement existing capability, safety, and compliance metrics, adding a clear, real‑world performance layer to today’s evaluation landscape.
\end{itemize}

This white paper describes FRAME's mission to help stakeholders better understand AI’s value in the real world, clarifies structural limitations in current evaluation approaches, and presents the centralized testing infrastructure for conducting large‑scale trials with structured, context‑rich observation. A concise glossary of key terms supports consistent language and shared understanding across the real-world AI evaluation ecosystem.

\section{Structural Limitations of the Current Ecosystem for Decision Making} 
AI use in the real world is not a set of isolated pass/fail tests, but an ongoing exchange in which human inputs and model outputs continuously shape one another and affect public life, including people who never touch AI directly.\cite{smith_can_2025,schulz-schaeffer_why_2025,zhou_attention_2025}. Deployments exhibit two interacting sources of variation:
\begin{compactitemize}
  \item \textbf{User entropy (the “host” variable):} The heterogeneity in how people phrase questions, adapt workflows, and interpret outputs; in the “epidemiology” of AI, this is the host environment.
  \item \textbf{Model stochasticity (the “agent” variable):} The inherent randomness in generative outputs. Small changes in decoding parameters or routing through mixture-of-experts layers can yield different responses to seemingly identical prompts, and this variability never fully disappears\cite{atil_non-determinism_2025}: this is the agent variable.
\end{compactitemize}

In the wild, user entropy and model stochasticity compound one another\cite{chen_prompt_2025,el-mhamdi_impossible_2022}: different users frame questions differently, and the model responds stochastically each time. The result is not a single, stable pattern of behavior but a moving distribution of interactions that can only be understood by modeling higher‑order effects over time, across many users and settings.  This turns AI deployments into wicked problems: complex socio‑technical interventions with distributed causes and effects and dynamically evolving feedback loops\cite{rittel_dilemmas_1973}. If evaluation is to serve deployment‑level decision‑making, it must move beyond certifying model performance to account for real-world variation, and generate evidence about how these tools are actually used, by whom, in which contexts, and with what consequences.

\subsection{A Missing Layer to \\ Support Sensemaking}
Research on expert–lay communication shows that a core bottleneck in real‑world decision-making is rarely a lack of technical detail. Instead, reliance on low‑context, highly technical outputs can \textit{increase} interpretive ambiguity, making it harder for decision‑makers to know what to do with the information\cite{corbin_how_2015,reyna_scientific_2021}. For example, accuracy metrics reported without context--such as a single benchmark score or “hallucination” rate--may appear precise but offer little guidance about what those numbers mean for specific users, tasks, or settings. What leaders and other decision‑makers often need most is sensemaking: the ability to integrate complex signals into simple, actionable “gists” about value and risk \cite{edelson_who_2024,fischhoff_risk_1993,reyna_theory_2008}. User entropy can serve as a central signal for sensemaking, since it captures how people actually appropriate, adapt, and interpret AI in the real world.

A public health analogy helps clarify what is missing from current evaluation practices. Today, most AI evaluation resembles a clinical trial, where lab‑style studies test a “compound” (the model) in tightly controlled environments to see whether it meets accuracy targets or predefined safety thresholds. Sensemaking, however, requires a layer of evidence more akin to epidemiology and post‑market surveillance. This additional layer can track what happens when AI systems are used in everyday settings, including side effects, adaptation, and longer‑term outcomes across diverse contexts\cite{schwartz_reality_2025}.

\subsection{Methodological Pitfalls: \\ AI Evaluation that Mirrors\\ Development}
Many evaluation methods overlook user entropy because they inherit the abstraction logic used in model training. Dominant approaches originate from developer‑centric practices that prioritize reproducibility and model improvement, but strip away the contextual detail that deployment‑level sensemaking depends on. By collapsing the user entropy side of real‑world deployment into a single “use case,” current approaches lock evaluation into sterile test sets and scripted in silico runs that capture only prompts and responses\cite{damour_underspecification_2020,liao_rethinking_2025}, and leave deployment‑facing questions under‑specified.

\begin{figure}[t]
  \centering
  \includegraphics[width=\linewidth]{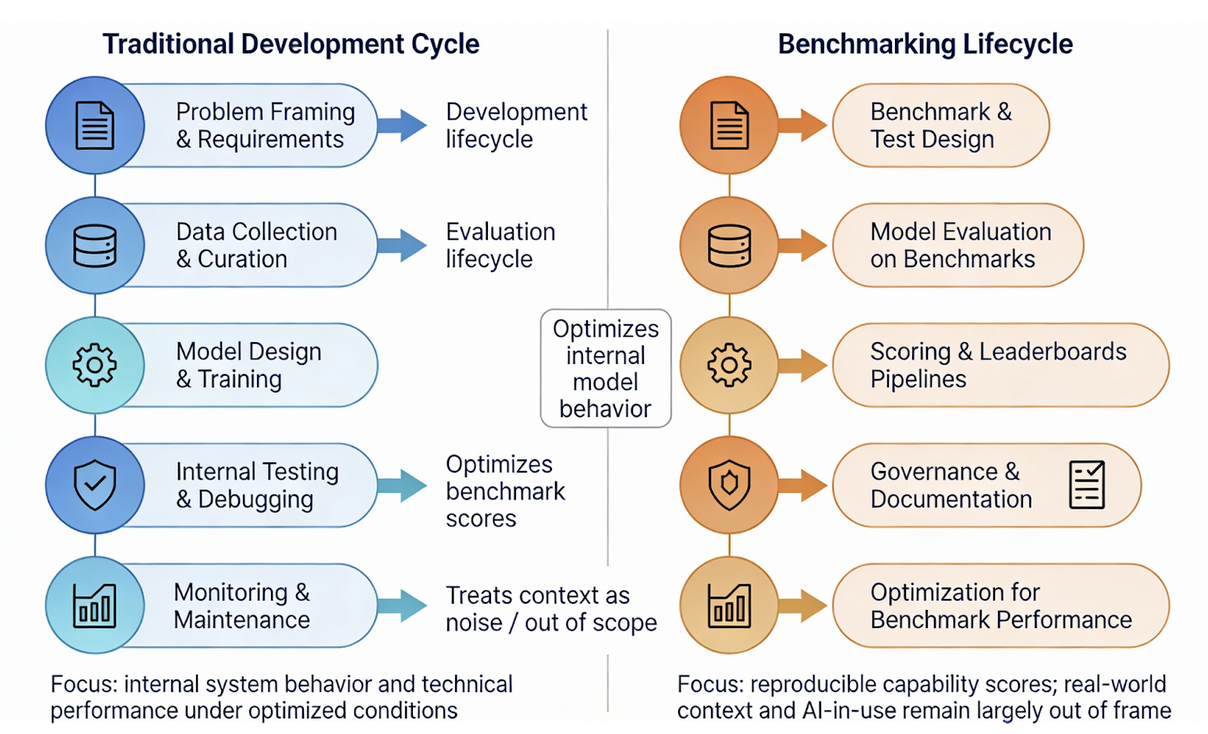}
  \caption{The current evaluation ecosystem uses methods that mirror the traditional AI development lifecycle, often neglecting user entropy and suppressing the context needed to make sense of outcomes for decision-making. (Generative artificial intelligence was used to support the creation of this graphic representing the authors’ own ideas, data, and words on this topic.)}
  \label{fig:evaluation-ecosystem}
\end{figure}

This decontextualization process also ends up rewarding performance in synthetic settings instead of everyday use, widening the gap between existing measures and the information needed to make sense of what happens across sites \cite{damour_underspecification_2020,wallach_position_2025}. As illustrated in Figure~\ref{fig:evaluation-ecosystem}, the current focus on model optimization leaves decision‑makers’ questions about deployment and implementation out of the measurement frame\cite{liao_rethinking_2025,bean_measuring_2025,salaudeen_measurement_2025}. Lacking evidence of AI's real-world outcomes, organizations struggle to judge whether deployments will generate real value in their own environments, and developers and risk teams must continually recalibrate guardrails and safety policies against shifting requirements \cite{barocas_fairness_nodate,boyarskaya_overcoming_2020}.

\paragraph{Partial Workarounds}
The current ecosystem relies on a set of strategies that serve as a substitute for direct, large‑scale evidence about what happens in practice. For example, harm assessments often use keyword‑matching safety filters that look for the presence of specific terms without regard for their role in context. A term such as “self‑harm” can appear in a public health awareness campaign, a news article, or promotional content for a mental health app or therapy service. A rigid, context‑blind safety metric will flag or block this benign content in the same way it would treat a user in an active mental health crisis or malicious content that encourages self‑harm.

These real‑world failures often stem from an inability to distinguish the mere presence of a term from its \textit{materialized} outcomes in context. Systems trained to rely on surface keywords show “shortcut” behavior: they over‑block harmless content and miss genuinely harmful content when keyword distributions shift\cite{CHEN2025114689,tasawong-etal-2025-shortcut}. This problem is exacerbated for higher‑order effects—such as psychological over‑reliance on AI—that build up gradually over thousands of everyday interactions. Other strategies exhibit similar structural issues:

\begin{itemize}
  \item \textbf{Alignment pipelines and preference data}: These approaches encode high‑level human values into models using “A versus B” forced‑choice labeling schemes tied to system guardrails and other stack‑level constraints. Alignment pipelines are central for tuning systems to specific norms but cover only a narrow band of possible behaviors, leaving many deployment‑relevant possibilities insufficiently specified or unmeasured, especially in heterogeneous settings\cite{atari_which_nodate,casper_open_2023,dahlgren_lindstrom_helpful_2025,conitzer_social_2024,weidinger_toward_2025}.
    \item \textbf{Technically scoped approximations}: Scripted agents and automated red‑teaming rely on hard‑coded interaction patterns and fixed prompt distributions\cite{wang_human-centered_2024} to surface specific bugs and toxic outputs. By removing real human variability, these tools can miss cumulative effects—such as changes in reliance on AI systems over time\cite{arhin2021groundtruthtruthexamining,chen_prompt_2025,meimandi_measurement_2025,plank_problem_2022,zhang_learning_2024,purpura_building_2025}.
    \item \textbf{Human judgment overlays}: These approaches leverage expert reviewers (e.g., lawyers) to label categories of outputs as acceptable or not. Because these judgments are made out of context rather than under real conditions, labels may not reliably predict real‑world outcomes or generalize across settings and populations\cite{bean_measuring_2025,salaudeen_measurement_2025,uher_rating_2022,crootof_humans_2023}.
    \item \textbf{Aggregation of capability evaluations}: Meta‑analyses pool benchmark scores across settings. By combining abstract tests without showing how systems actually behave in context, these approaches offer limited support for sensemaking and can compound underlying sampling errors from in silico testing\cite{jiang_beyond_2026,eriksson_can_2025,schwartz_reality_2025}. 
    \item \textbf{Governance and compliance tools}: Risk taxonomies, compliance checklists, and explainability dashboards demonstrate alignment with technical or policy requirements, and are well‑suited to system design, documentation, and assurance. To fully support deployment-focused sensemaking and decisions, these tools also need details about how systems behave with real users in real contexts.\cite{eriksson_can_2025,schwartz_reality_2025,ntia_accountability_2024}.
    \item \textbf{Isolated context‑aware studies}: Context‑aware methods—such as usability tests, adoption studies, and domain‑specific pilots—offer valuable local insight into how people interact with systems\cite{berman_scoping_2024}. Yet they rarely use a shared structure for collecting and organizing findings, making it hard to compare patterns across sites or connect specific interactions to broader secondary and tertiary effects\cite{matias_humans_2023,meimandi_measurement_2025,weidinger_sociotechnical_2023,khullar_nurturing_2025,punton_reality_2020,Ibrahim2025InteractiveEvaluations}.
\end{itemize}

The strategies listed above generate signals that feed back into model development to make systems more usable and safer, but they are often too localized, partial, or inconsistent to support deployment‑level decision‑making.\cite{damour_underspecification_2020} This reflects a structural weakness: existing tools can track patterns in inputs and outputs, but because they struggle to capture the dynamics, feedback loops, and behavioral adaptations that shape AI‑in‑use, they cannot reliably support deployment‑level requirements.

Some recurring gaps illustrate why it is so hard to form a coherent picture across measurement strategies and their outcomes:
\begin{itemize}
    \item \textbf{Obscuring operational friction}: Model‑centric success metrics can hide everyday friction points; a “correct” output may still require substantial human verification or rework, turning a theoretical efficiency gain into an operational burden\cite{bean_measuring_2025,eriksson_can_2025,liao_ai_2023,zhuang_position_2025}.
    \item \textbf{Ignoring downstream consequences}: When safety evaluations treat “hallucinations” only as abstract factual errors, organizations lose sight of what happens next. For example, do users notice and correct errors or instead trust and propagate them into other materials \cite{ashktorab_emerging_2024,suzgun_language_2025,macnamara_does_2024,meimandi_measurement_2025}.
    \item \textbf{Scores with no context}: Aggregate figures—such as an “11\% hallucination rate” or a single safety score—rarely indicate 11\% of what, for whom, or under which conditions, and models that look similar on these metrics can behave very differently across subpopulations and settings \cite{damour_underspecification_2020}.
    \item \textbf{Masking contextual harms}: Benchmarks tuned for narrow accuracy are poorly suited for generative systems, where there is no single “right” paragraph, image, or audio clip. Systems can easily be factually correct yet still harmful. For example, a public‑service chatbot can produce accurate information in a format that is inaccessible to people with disabilities\cite{lebovitz,eriksson_can_2025,panda_accesseval_nodate}.
\end{itemize}

\subsection{AI Evidence Layers}
The ecosystem limitations described above highlight the need for connective tissue between existing methods. Systematic, deployment‑facing knowledge must link abstract evaluations to contextual insight, so evidence can travel across sites while remaining grounded in real use.  To address this need, we distinguish three layers of evidence.

\begin{figure*}[tbp]
  \centering
  \includegraphics[width=0.8\textwidth]{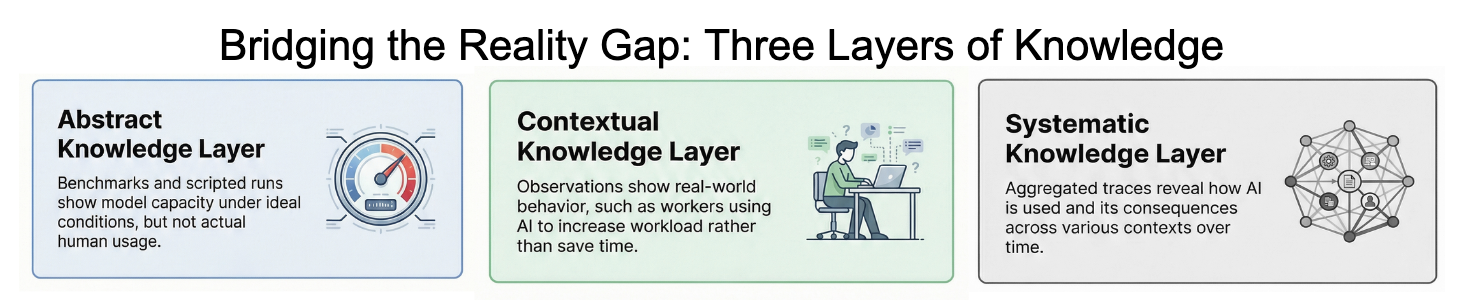}
  \caption{An example of how three knowledge layers build up evidence across contexts to address the decision‑maker's dilemma.
 (Generative artificial intelligence was used to support the creation of this graphic representing the authors’ own ideas, data, and words on this topic.)}
  \label{fig:layers}
\end{figure*}

\paragraph{Layer 1: Abstract knowledge}
Benchmarks, alignment pipelines, and related tools produce abstract knowledge about what a model can do in principle\cite{maslej2025artificialintelligenceindexreport,eriksson_can_2025}. Like clinical lab results, these tests assess primary effects and answer development‑focused questions, such as:
\begin{itemize}
  \item “Can the model solve algebraic equations on this standardized test set?”
  \item “Did the model resist adversarial attacks during automated red‑teaming?”
  \item “Did it score higher on ‘helpfulness’ in generic preference rankings?”
\end{itemize}

\paragraph{Layer 2: Contextual knowledge}
Methods such as user research, audits, domain‑specific pilots, red‑teaming, and incident reports produce contextual knowledge about outcomes in particular settings \cite{berman_scoping_2024,deng_understanding_2023,dwyer_mindbenchai_2025,khullar_nurturing_2025}. Analogous to medical case reports, they surface secondary effects—near‑term impacts such as workflow shifts or harms—and answer questions like:
\begin{itemize}[itemsep=0pt, topsep=2pt, parsep=0pt, partopsep=0pt]
  \item “Which responses did users prefer in this usability test?”
  \item “Are harms or unintended consequences occurring in this deployment?”
  \item “What failures or workarounds were logged during this pilot?”
\end{itemize}

\paragraph{Layer 3: Systematic knowledge}
Longitudinal field studies, multi‑site monitoring, and other real‑world testing strategies generate systematic knowledge about cumulative impacts across organizations and over time\cite{meimandi_measurement_2025,schwartz_reality_2025,weidinger_sociotechnical_2023}. Like epidemiological surveillance, these approaches can show how AI is used, by whom, in which situations, and with what consequences. By bringing scalable testing into real settings and combining it with structured observation, user entropy can be treated as signal rather than noise. Resulting evidence from such tests can help decision‑makers answer questions such as:
\begin{itemize}[itemsep=0pt, topsep=2pt, parsep=0pt, partopsep=0pt]
  \item “Is this tool creating rework for my staff, or actually saving time?”
  \item “Are employees over‑relying on the tool and losing critical skills?”
  \item “Is the tool shifting liability to our frontline workers?”
  \item “Who is consistently benefiting, and who is absorbing new risks?”
\end{itemize}

A recent field study of generative AI in the workplace illustrates these layers in practice \cite{ye_ai_2026}. When only Layer 1 abstract metrics about generative AI capabilities are available, organizations might presume productivity gains. The workplace study used observational approaches to corroborate those presumptions, generating Layer 2 contextual insights that revealed added work and hidden friction for employees instead of freeing up time. Workers juggled multiple AI‑mediated tasks, expanded their job scope, and experienced more intense, fragmented work. This "contextual knowledge" offers leaders the kind of evidence they need to manage AI use locally \cite{ye_ai_2026}.

Real‑world AI methods can build on such studies to generate systematic knowledge (Layer 3). These evaluations can link system behavior, the context in which it operates, and real‑world outcomes at scale to surface higher‑order effects that emerge over time. They build evidence to support both organizational decision‑making and cross‑sector claims \cite{schwartz_reality_2025}. Figure~\ref{fig:layers} shows these layers, using the workplace study as an example.

\paragraph{Analogies from Consumer Technology}
Systematic, cross‑context measures have been used to understand the real‑world impacts of other technologies:
\begin{itemize}
\item \textbf{Smartphone telemetry}: Large‑scale telemetry and screen‑time data revealed interaction patterns that individual user studies missed, linking intensive smartphone use to outcomes such as sleep quality and attentional performance\cite{duke_smartphone_2017,christodoulou_phone_2025}. This evidence informed features like “do not disturb” defaults and digital best‑practice guidelines\cite{skowronek_mere_2023}.
\item \textbf{Traffic and GPS data}: Aggregated GPS traces turn individual trips into measures of route reliability and travel‑time variability\cite{raez_inrix_2026,woodard_predicting_2017,elefteriadou_highway_2016}, helping travelers choose when and where to travel.\cite{liu_analysis_2022,klaver_data_2025,wang_empirical_2020}
\end{itemize}

\section{How FRAME Works}
FRAME is designed to complement the current ecosystem by modeling AI variation in the real world and produce structured, descriptive evidence about how systems are actually used and the outcomes they produce in context. Because current approaches are unable to reliably simulate such behavior, FRAME aims to collect it at scale via systematic observation of AI-in-use without stripping away real-world variability. While traditional human‑subject testing is often viewed as slow and manual, FRAME's centralized infrastructure and "methods lab" enables speed and reproducibility. 

\begin{table*}[t]
\centering
\renewcommand{\arraystretch}{1.2}
\begin{tabular}{|P{0.20\textwidth}|P{0.33\textwidth}|P{0.33\textwidth}|}
\hline
& \parbox[t]{0.33\textwidth}{\centering \textbf{Contextual scoring}\newline\textbf{(human-grounded)}}
& \parbox[t]{0.33\textwidth}{\centering \textbf{LLM-as-a-judge}\newline\textbf{(automated)}} \\ \hline
\textbf{Panelist traces}\newline\textbf{(user entropy)} &
Human participants work through realistic scenarios; their traces and reports are organized via contextual rubrics into descriptive categories of use, non-use, reliance, and abandonment, making patterns of use visible across contexts. &
The same participant outputs are also labeled by an LLM using the shared rubrics, producing parallel descriptive segmentations that show how the model summarizes the friction, workarounds, and value that people report. \\ \hline
\textbf{Scripted chatbot runs}\newline\textbf{(abstract knowledge)} &
Scripted chatbot runs on the same scenarios are categorized with the contextual rubrics to describe how the system behaves under controlled conditions, separate from human behavior. &
Scripted runs are also labeled by an LLM, creating a reusable, fully automated descriptive pathway for rapid comparison and regression-style checks across models and configurations. \\ \hline
\end{tabular}
\caption{Complementary human-grounded and automated descriptive scoring across panelist traces and scripted chatbot runs help reveal the gap between automated and real-world perspectives of sandbox activity.}
\label{scoringruns}
\end{table*} 

\subsection{Testing sandbox}
FRAME’s testing sandbox is a controlled yet realistic environment for sponsor‑led, scalable trials of commercial systems across domains such as education, media, and consumer services. It is instrumented to capture system behavior under two parallel streams, with shared test scenarios, detailed logging, and common scoring rubrics to connect them.
\begin{enumerate}
  \setlength{\itemsep}{0pt}
  \setlength{\parsep}{0pt}
  \setlength{\parskip}{0pt}
  \setlength{\topsep}{0pt}
  \setlength{\partopsep}{0pt}
  \item \textbf{Remote Participant Panels:} a fully dynamic evaluation stream to observe how thousands of people actually use AI in realistic, scenario-based tasks. Panelists complete structured, task‑based scenarios and report what actually occurred, including how they adapted, repurposed, or worked around system outputs. Individual participants are neither identified nor directly measured. Panelists annotate their own interactions using a common response protocol, capturing how requests are phrased, when confusion arises, and when they give up or switch tools. To minimize participant risk, the sandbox uses proxy scenarios and standard human-subjects protections.
  \item \textbf{Scripted Chatbot Runs:} a fully automated, in‑silico counterpart that runs the same scenarios as the panels through scripted chatbot sessions to reveal the model’s ideal performance under controlled conditions.
\end{enumerate}

\noindent\textbf{Test scenarios} The mechanism by which FRAME sets evaluation task parameters, scenarios support repeatable testing across systems, configurations, and user groups. Panelists are free to interact with AI systems as they normally would. FRAME logs telemetry for each session, tracking factors such as whether participants used or ignored the AI, how many steps they took, if they overrode recommendations, and whether they reverted to old workflows. Over time, these traces build a reusable corpus of AI‑in‑use that feeds FRAME’s community models--digital summaries of how evaluated systems were used in the sandbox. Organizations and policymakers can use community models to test policies and compare systems across sites without building full evaluation pipelines from scratch. enabling comparisons across deployments and informing sector‑specific baselines.

\textbf{Scoring}: FRAME uses a dual‑stream scoring engine in which the object of evaluation is the AI system, not the people using it. Panelist traces and scripted chatbot runs are both labeled with scenario‑specific rubrics to capture descriptive detail about how systems are used, ignored, or abandoned in context. An LLM‑as‑judge applies the same rubrics automatically to generate parallel descriptive codes and scores for system behavior. 

Comparing scored outputs can highlight differences between automated runs and human-grounded approaches for surfacing details about AI's value, friction, and risk. Table~\ref{scoringruns} summarizes the kinds of insights that can be surfaced by applying human‑grounded and LLM-as-a-judge scoring to the chatbot and panelist runs. 

{\setlength{\parskip}{0.5\baselineskip}
\subsubsection{Illustrative example}
These insights can help address a common organizational question when deploying AI: which workflows can be fully automated, and where, when, and how is human expertise still necessary to provide local, contextual, or domain‑specific knowledge? For example, transportation‑sector decision‑makers might want to determine which steps in routing, exception handling, and customer notification can safely run end‑to‑end on AI, and where dispatchers or operators still need to review, correct, or supplement system outputs before they affect passengers or freight. 

Sponsors from the transportation sector might work with FRAME members to design realistic evaluation scenarios for deployment to the sandbox. Participant panels can use either specialized tasks tailored to transportation experts or more general tasks that any panelist can complete, with high‑volume automated trials run in parallel on the same tasks. By comparing chatbot runs and panelist traces, differences between scripted and real‑world behavior --including use, non‑use, reliance, and abandonment--become visible, providing stakeholders with clarity about the kinds of settings and system–user configurations that are suitable for their contexts.

\begin{figure*}[tbp]
  \centering
  \includegraphics[width=0.8\textwidth]{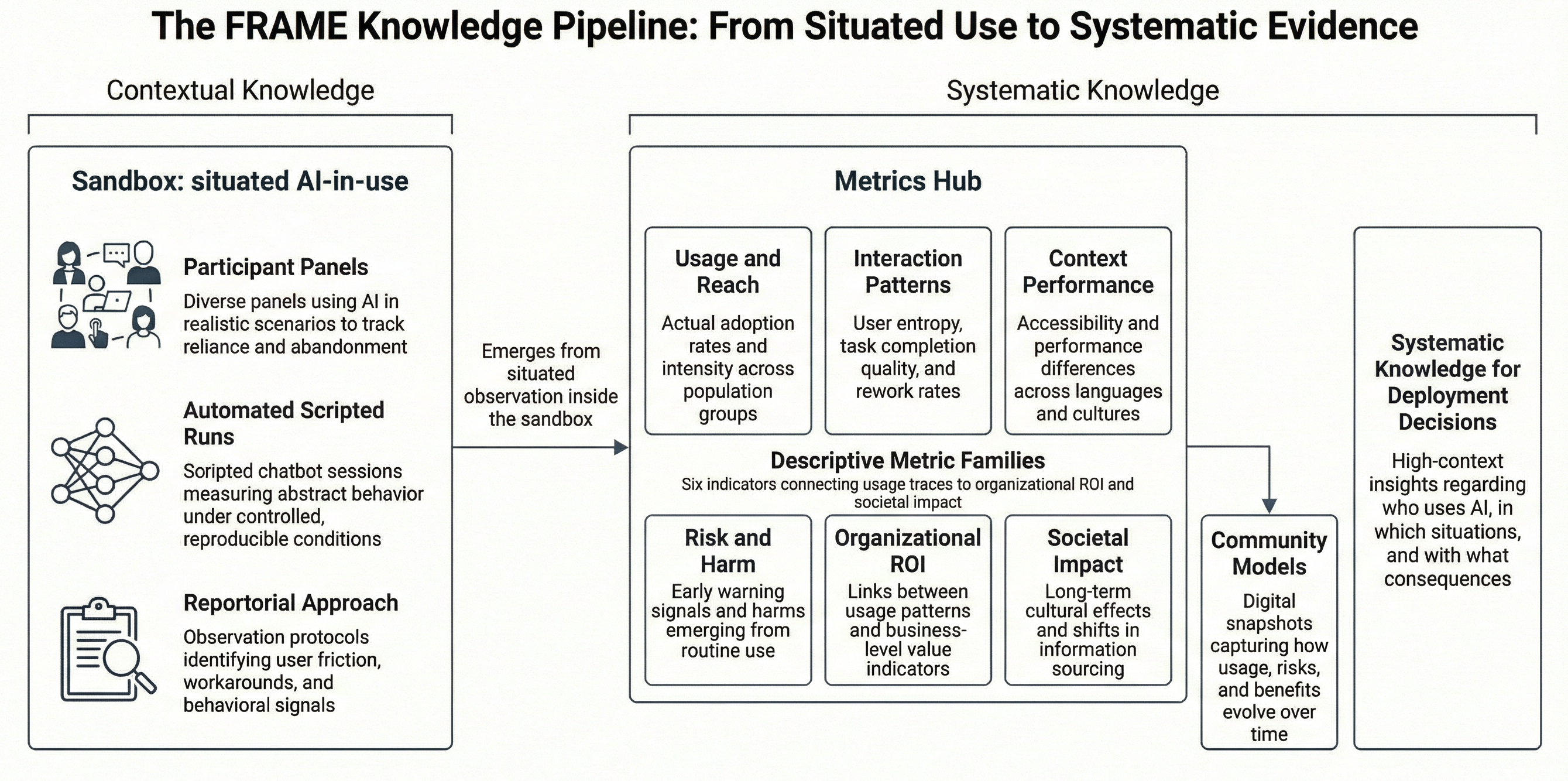}
  \caption{Overview of FRAME’s testing sandbox, metrics hub, and community model architecture. All measures are derived from instrumented sandbox interactions, where panel participants work through realistic scenarios under built‑in safety protections. (Generative artificial intelligence was used to support the creation of this graphic representing the authors’ own ideas, data, and words on this topic.)}
  \label{fig:frame-architecture}
\end{figure*}

}
\subsection{Metrics Hub}
Basic questions about AI use are still hard to answer. Many organizations must rely on proprietary dashboards, one-off surveys, or self-reported pilots and cannot say with precision how many people actively use AI tools in their settings, which tools they rely on, how use varies across roles and communities, or whether some groups are consistently routed through lower-performing models. Internal telemetry can show logins or API calls, but it rarely shows how tools are embedded in tasks, when people abandon them, or how usage connects to outcomes like access, quality, safety, utility, or impact. 

FRAME’s metrics hub treats these questions as descriptive measurement targets in their own right. Each evaluation reports out which groups used the systems, in which settings, what happened during and after the interaction, and with what results. Indicators illuminate patterns of use, non‑use, workarounds, and reliance across groups and configurations. They are designed to sit alongside existing capability, safety, and compliance metrics, adding a deployment‑facing layer rather than replacing current tools. FRAME metrics are generated outside vendor reporting pipelines, using protocols and standing panels governed by academic and community‑oriented standards. This external vantage point gives leaders actionable, context‑specific insight into how systems behave in practice to corroborate, challenge, or extend company claims.

 Figure~\ref{fig:frame-architecture} illustrates how FRAME's sandbox and metrics hub can support a practical systematic knowledge base for deployment decisions across settings.

Selected metrics are released on a regular cadence to complement model leaderboards—allowing evaluators to validate or challenge benchmark claims. The hub maintains a focus on inquiry and learning rather than new optimization targets, aligning with recent critiques of leaderboard‑centric evaluation and reinforcing a shift toward more context‑aware, deployment‑focused assessment \cite{thomas_reliance_2022,singh_survey_2025}. 
\paragraph{Metrics families}
\label{metricsfamilies}
To organize these insights, the metrics hub groups sandbox outcomes into six families of decision‑ready indicators that can be compared across sectors and deployments. Some headline indicators from each family will be released through public‑facing dashboards, while more detailed views remain configurable for specific sponsors and use cases. In each family, specific indicators will vary by sector, sponsor questions, and available data. the examples below are not a fixed checklist, but illustrate the kinds of measures FRAME prioritizes

\begin{enumerate}[itemsep=4pt, topsep=2pt, parsep=0pt, partopsep=0pt]
  \item \textbf{System usage and reach} Describes actual system adoption and participation. Examples may include, active-use rates of systems across the sandbox, intensity and frequency of use, clustering by task type, and patterns of use and non-use across groups, contexts, and configurations (where consent and protections permit).
  \item \textbf{Task and interaction patterns} Characterizes whether and how sandbox participants integrate the systems into their day and how user entropy shows up in practice. This may include: error and rework rates, uplift versus added burden, workarounds, repurposing, misuse and over-reliance, cognitive load from checking and fixing AI output, and how behaviors like sycophancy and anthropomorphization build over time for different user groups.

  \item \textbf{Access and performance across contexts} Describes how system behavior differs across sandbox participant groups and scenarios. This may include: compatibility with assistive technologies,\cite{waters_aitesting_2026} readability and perceived accessibility of outputs, success rates across languages and dialects, and extra work required by specific population groups to attain comparable outcomes.

  \item \textbf{Risk and harm signals in routine use} Tracks "early‑warning signs" in sandbox scenarios that approximate unsafe or problematic interaction. Examples may include the conditions under which proxies for misleading, harmful, or confusing outputs appear in panelist interactions; the time and effort the panelists needed to monitor and correct them; and proxy indicators of heavy or “binge‑like” use that may shift panelist information diets over time.

  \item \textbf{Organizational value and ROI} Links panelist usage patterns in FRAME‑designed scenarios to existing organizational indicators for sponsor-informed use cases. This may include how panelists used AI-mediated workflows, changes in decision speed, rework, appeals, and complaints.

  \item \textbf{Downstream cultural and societal impact} Surfaces longer‑term, higher‑order effects on culture and communities through repeated waves of sandbox studies over time. This may include changes in how panel participants find and share information; how panelists align with or defer to the system over time; and shifting norms around language use.
\end{enumerate}
\begin{table*}[t]
\centering
\caption{Decision lenses, reference standards, and guiding questions in the AI ecosystem.}
\label{tab:decision-lenses}
\small
\setlength{\tabcolsep}{4pt}      
\renewcommand{\arraystretch}{1.1} 
\begin{tabular}{p{4cm} p{6cm} p{6cm}}
\hline
\textbf{Decision lens} & \textbf{Reference standards for performance} & \textbf{Guiding question} \\
\hline
Real-world \newline evaluation (FRAME) 
& Systematic evidence about AI-in-use, including user entropy and higher-order effects 
& What happens when people actually use the tool in this context (does it create value or new risks/burdens)? \\
\hline
Model / research (ML) 
& Labeled data, benchmarks, leaderboards 
& Did the model get the right answer on this test set? \\
\hline
Product engineering 
& Customer requirements, feature usage, tickets 
& Does the system solve the problem well enough to ship? \\
\hline
Compliance 
& Policies, laws, internal risk registers 
& Did we break a specific rule? \\
\hline
UX and usability 
& User feedback, adoption and retention metrics 
& Can people use the interface as designed? \\
\hline
Safety and alignment 
& Human preference data, preference labels, safety taxonomies, red-teaming results 
& Did the model pass the safety stress test and produce policy-compliant outputs? \\
\hline
\end{tabular}
\end{table*}

\subsection{Methods and Insights Lab}
Within the distributed consortium anchored at Virginia State University’s Center for Responsible AI, FRAME members draw on technical, social, organizational, and lived expertise to: 
\begin{itemize}[itemsep=0pt, topsep=2pt, parsep=0pt, partopsep=0pt]
  \item Build a shared infrastructure and methodological backbone to systematically observe AI‑in‑use across many contexts and build up a cumulative evidence base.
  \item Formalize how interactions and outcomes are logged, to capture where AI fits within workflows, when and why people stop using it, and what happens next (rework, delay, reliance, burden shifting).
  \item Turn sandbox findings into clear, decision-ready insights that help organizations make sense of AI’s value and risks in their own settings.
\end{itemize}

\section{Reorienting AI Evaluation\\ Toward an Evidence Layer \\ for Decision-Making}
Industry leaders increasingly acknowledge that traditional capability benchmarks—though still necessary—are no longer sufficient on their own. As Microsoft CEO Satya Nadella recently noted,\footnote{https://www.moneycontrol.com/technology/microsoft-ceo-satya-nadella-says-ai-needs-to-help-healthcare-education-and-productivity-not-just-burn-energy-article-13190215.html} “the real benchmark for AI progress is whether it makes a real difference in people’s lives—in healthcare, education, and productivity,” rather than how well models perform on challenge tasks. FRAME’s contribution is to help build an evidence layer that links model‑centric metrics to these deployment‑level questions about value, risk, and burden in real settings.

\subsection{Formalizing Systematic Knowledge \\ for AI }
Reliable insights into where AI will deliver value—or what outcomes to expect in practice—remain scarce. Even organizations well along the adoption curve encounter friction when moving from pilot projects to sustained use \cite{challapally_genai_divide_2025}. From FRAME’s perspective, the continuing priority on model performance leaves decision‑makers with abstract results, ad hoc experiments, and reactive problem‑solving—rather than grounded guidance about everyday use\cite{mckinsey,ransbotham_emerging_2025}.

While many organizations may hold both abstract model‑centric metrics and rich contextual knowledge about AI‑in‑use, they typically lack structured ways to turn that information into evidence that can guide institutional decisions\cite{weidinger_sociotechnical_2023}. The field of risk communication emerged from a similar challenge in public health, as raw epidemiological data and specialist reports proved insufficient for policymaking\cite{saracci_introducing_2015,windle_epidemiologic_2019,calleja_public_2021,piret_pandemics_2021}. As a distinct practice, risk communication focuses on translating complex, ambiguous evidence into usable guidance for decision‑makers and the public\cite{bostrom_vaccine_nodate,fischhoff_risk_1993,balogway_evolving_2020}. Real‑world AI evaluation can play an analogous role: not a replacement for capability or benchmark testing, but a systematic knowledge layer that links those measurements to the practical choices facing deployment stakeholders. 

Many organizational policies require Testing, Evaluation, Verification, and Validation (TEVV) as a foundation for trustworthy AI, yet few can produce evidence at the deployment level\cite{flournoy_building_2020,maslej2025artificialintelligenceindexreport}. FRAME’s real‑world evaluation lens can help organizations operationalize these frameworks—particularly the “T” (Testing) and the second “V” (Validation) by translating high‑level TEVV requirements into concrete evidence of how systems function in real‑world contexts\cite{schwartz_reality_2025,chouldechova_shared_2024} that existing safety, alignment, and compliance tools often overlook. Grounding evaluation in real events can also enable the production of higher‑quality datasets for training and governance \cite{Ibrahim2025InteractiveEvaluations,Kaptein2017UncoveringNS, larsen_validity_2025}, reducing dependence on canned user personas and on inferring behavior solely from the extreme ends of carefully curated test conditions \cite{collins_evaluating_2023,harris_forgotten_2025}.

\subsection{Decision Lenses Across \\ the AI Ecosystem}
The AI ecosystem is not a single, unified structure, and success looks different depending on who is asking the question. For model developers, success typically means improving system performance. For engineers, it revolves around meeting customer needs and technical requirements. Real‑world evaluation adds a third lens—one centered on deployment decisions—treating evidence about how AI is actually used as the primary reference point for judging success. Table~\ref{tab:decision-lenses} shows how this real‑world lens differs from other organizational functions.

\subsection{Three Shifts for Real-World \\ AI Evaluation Practice}
\label{operationalizingshifts}
To build up this real-world lens FRAME's ecosystem shifts who evaluation serves, what it measures, and how it is conducted. 

\subsubsection{Who Evaluations Are Built For: \\ Decision-making Beyond the Stack} 
Generative systems are more than software. They function as products that people use and information media they interpret, trust, or ignore \cite{capraro_impact_2024,wang_human-centered_2024,smith_can_2025,schulz-schaeffer_why_2025}. The same underlying model can be used as a writing assistant, a search interface, or an advice source, often within a single workflow \cite{tamkin_clio_2024,kehoe_top_nodate}. Evaluation infrastructures tend to mirror the audiences they are built to serve, and today those audiences are mostly inside the stack—model developers and product teams—rather than the people and institutions closest to deployment. As a result, this variability in how systems are used, governed, and incentivized is often left out of view. FRAME responds by expanding evaluation to serve decision‑makers beyond the stack, treating them as a distinct audience with its own evidence needs.

\paragraph{AI Users as Reporters}
With a primary focus on model training, most benchmarking regimes effectively treat people as measuring instruments: they check whether a model stayed within preset bounds and label or rank outputs based on static preferences (for example, “Response A is better than Response B”) to optimize accuracy or preference alignment on a fixed target. Systems are tuned to that endpoint even when it does not reflect real use\cite{bean_measuring_2025,wallach_position_2025,liao_rethinking_2025,salaudeen_measurement_2025}. 

Yet, accuracy serves as only one part of the equation when aiming to understand real-world behavior. For example, evaluators need to do more than identify hallucinated output. Determining the real impact of hallucinations requires knowing what people do with that content—whether they notice it, believe it, correct it, or propagate it into later decisions--and how these choices accumulate into broader outcomes such as changes in skill and judgment \cite{selgas-cors_sociotechnical_2025,xu_generative_2025,xiao_ai_2025,agrawal_ai_2021}. While more comprehensive information about real-world use is often captured via platform mechanisms such as feedback buttons or incident logs, it can serve as a lagging indicator and be prone to survivorship bias.  

To address these measurement challenges, panel participants in FRAME's sandbox act as observers and reporters rather than raters, performing AI‑mediated tasks and documenting what actually happens. This can include detail about where panelists may improve, adapt, or repurpose AI outputs, quietly work around failures or disengage altogether \cite{collins_evaluating_2023,harris_forgotten_2025}. The sandbox leverages descriptivist methods to identify such  patterns amid individual variation \cite{gelman_hennig_2017,Kaptein2017UncoveringNS,biber_representativeness_1993,verbeke_ecological_2024}. In aggregate, panel reports form a descriptive corpus of interaction in which overlapping perspectives yield a scalable, contextual view to reveal:
\begin{itemize}
    \item How systems behave, where they create friction, and who is most affected.
    \item How people integrate, adapt, or ignore system outputs in real workflows.
    \item Where friction, workarounds, or new risks emerge in everyday use.
    \item Which outcomes matter most to different stakeholder groups, and how AI reshapes those outcomes over time.
\end{itemize}
\begin{figure}[tbp]
  \centering
  \includegraphics[width=0.9\linewidth]{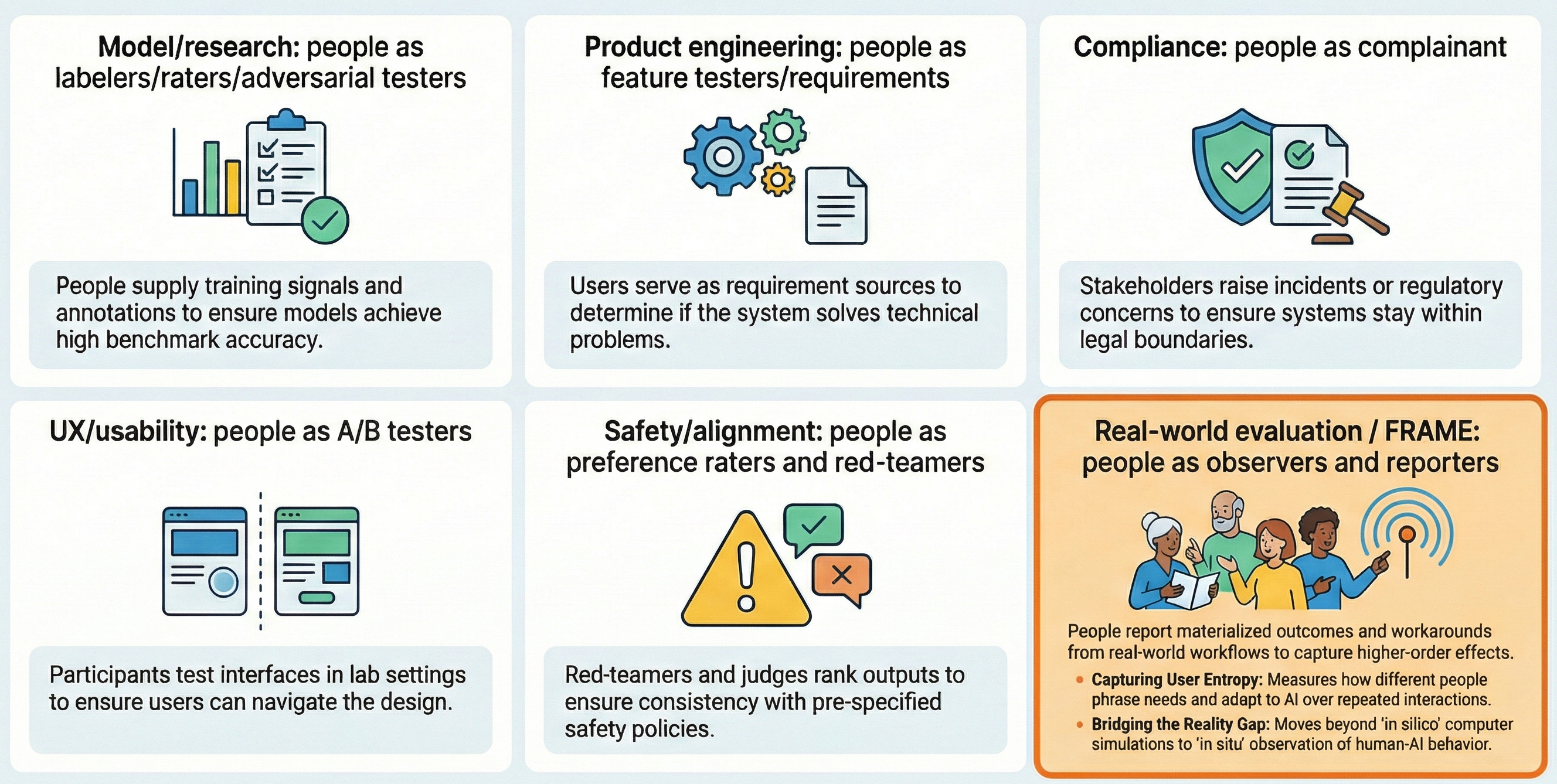}
  \caption{How different decision lenses position people in AI evaluation.(Generative artificial intelligence was used to support the creation of this graphic representing the authors’ own ideas, data, and words on this topic.)}
  \label{fig:reporters}
\end{figure}
As shown in Table~\ref{tab:validity}, foregrounding the questions decision‑makers actually face--and shifting evaluation methods accordingly-- can also enhance key forms of validity. 

\begin{table*}[t]
\centering
\caption{How types of validity adapt between in-silico and in situ evaluations.}
\label{tab:validity}
\setlength{\tabcolsep}{4pt}      
\renewcommand{\arraystretch}{1.1} 
\small
\begin{tabular}{p{4cm} p{6cm} p{6cm}}
\hline
\textbf{Validity type} & \textbf{Model-centric (in-silico)} & \textbf{Real-world evaluation-FRAME (in situ)} \\
\hline
Content and construct (Are we measuring the right thing?) 
& Weakened by reliance on abstract proxies (such as perplexity or generic preference scores) that miss how people actually express needs in deployment\cite{larsen_validity_2025}.
& Strengthened through metrics based on observed user behavior and stakeholder‑defined goals\cite{salaudeen_measurement_2025}. \\
\hline
Ecological (Does the test \newline resemble the real‑world use conditions?)
& Undermined by dependence on “sterile” prompt‑response tests that do not reflect real workflows.
& Strengthened through realistic scenarios and situated observation that capture friction, workarounds, and constraints. \\
\hline
Consequential (What happens downstream\newline when these tools are used?)
& Underdeveloped by failing to capture how outputs alter real‑world outcomes over time.
& Strengthened by tracking whether AI use creates value or shifts burden. \\
\hline
External (Will the test results hold for my setting?)
& Uncertain: Lessons from one optimized test set often fail to generalize to different users and contexts over time.\cite{campbell_experimental_1959}
& Strengthened by testing across diverse groups and settings at scale. \\
\hline
\end{tabular}
\end{table*}

\subsubsection{What Evaluations Target: From Accuracy to Higher-Order Effects}
To support increasing interest in the higher‑order effects of system use\cite{farrell_large_2025,reuel_betterbench_2024}, FRAME's sandbox is instrumented to track how impacts accumulate for panelists across repeated interactions and scenario-based tasks over time. FRAME’s approach makes these dynamics observable by jointly modeling user entropy and model stochasticity, capturing how variation on both sides shapes real‑world performance and surfaces higher‑order effects. These outcomes can complement traditional capability benchmarks and inform how synthetic agents are tuned so their prompts, task choices, and abandonment rates more closely reflect real‑world behavior \cite{li_hugagent_2025}.

\begin{table*}[t]
\centering
\renewcommand{\arraystretch}{1.2}
\begin{tabular}{|l|p{0.35\textwidth}|p{0.35\textwidth}|}
\hline
\textbf{Dimension} & \textbf{Status-quo AI Evaluation} & \textbf{Real-world AI Evaluation (FRAME)} \\ \hline

\textbf{Who} &
\begin{tabular}[t]{@{}l@{}}
Model builders \\
\textit{(inside the stack processes)}
\end{tabular} &
\begin{tabular}[t]{@{}l@{}}
Deployment stakeholders \\
\textit{(outside the stack decision making)}
\end{tabular} \\ \hline

\textbf{What} &
Isolated model capabilities &
Downstream impacts and real-world variability \\ \hline

\textbf{How} &
\begin{tabular}[t]{@{}l@{}}
Prescriptive model-centric checks \\
\textit{(AI governance intent)}
\end{tabular} &
\begin{tabular}[t]{@{}l@{}}
Observing AI at scale in the wild \\
\textit{(AI governance reality)}
\end{tabular} \\ \hline

\end{tabular}
\caption{How FRAME's real-world AI approach shifts evaluation practice to support deployment decision-making.}
\label{tab:statusquo-vs-frame}
\end{table*}

\subsubsection{How Evaluation is Done: \\ From Static Tests to \\ Ongoing Observation} 
Current evaluation approaches often serve primarily as compliance checks of “governance intent” (“was the rule followed?”) Real‑world AI evaluation instead generates higher‑resolution evidence of AI‑in‑use, such as whether a tool saves time, adds friction, or shifts burden. Through descriptive tracking of what \textit{actually} happens as panelists work through sandbox scenarios (e.g., "did it save time?"), FRAME can surface "governance reality" and enable insights for improved decision making.

This shift also targets a core friction in decision‑making: ambiguity. High‑context, low‑ambiguity information about who is using AI, where, for what, and when helps stakeholders interpret findings against their own goals\cite{zakaria_edward_2016}. Low-context benchmarks, by contrast, create ambiguity: the same efficiency or accuracy number can hide wide differences across contexts and user groups \cite{park_we_2024}, making it hard to know what those results mean for any specific deployment.

With dedicated infrastructure, standing panels, and shared scenarios, real‑world tests can run on a regular schedule, similar to large field surveys of public opinion. Over time, sector‑specific "digital twins" of users, tasks, and environments can help automate evaluation by comparing new system behavior to baselines grounded in prior human data.

\section{Limitations and \newline Operational Boundaries}
Decision‑makers need more than leaderboards; they need ways to navigate real deployments. FRAME evaluates AI‑in‑use and turns outcomes into systematic knowledge so organizations can move from guessing about AI’s impact to actively managing it—deciding where to invest, what to scale, and how to protect their operations and communities. At the same time, producing decision‑ready evidence introduces its own complexities, trade‑offs, and practical limits, some of which are outlined below.

\textbf{Balancing human observation with automated scaling}
FRAME uses automated tools while keeping people at the center of evaluation so insights can scale across scenarios, systems, and configurations. Traditional in‑silico testing is indispensable for probing model behavior and generating abstract knowledge, but on its own it cannot show how AI systems actually work in people’s lives. The sandbox still uses in‑silico tests, but pairs them with structured human observation to produce deployment‑level insights about the strengths and limits of each approach.

FRAME’s digital twins serve as another automated scaling technique: they use observed behavioral traces to calibrate simulations so they reflect actual user entropy rather than purely hypothetical users. LLM‑as‑judge scoring is used as an automated way to produce descriptive baselines, not as a source of ground truth. Its outcomes are compared against human‑grounded results to reveal key gaps between automated and dynamic scoring methods.

\textbf{Using proxies without losing the plot}
Proxies are often used in AI evaluation to stand in for richer real‑world concepts that are hard to observe directly, and their value depends on how well they are grounded in context.\cite{bucinca_proxy_2020} In FRAME’s real‑world methods, carefully chosen proxies help approximate complex constructs closely enough to make measurement practical while also reducing exposure to harm and protecting sensitive data. For example, severe emotional dependence on AI systems cannot be ethically studied directly, but proxy scenarios can model underlying mechanisms under realistic conditions and with strong guardrails.

When proxies are vague, weakly justified, or never validated against the constructs they are meant to represent, they can become misleading shortcuts that distort what “good” performance looks like. For example, “passing a safety filter” can serve as a stand‑in for “safe behavior in the wild”. When disconnected from real-world use, such proxies can shift evaluation toward what is easy to count (such as keyword refusals) rather than what matters for decision‑making.

\textbf{Centralizing the resource burden of real‑world evaluation}
Running situated observation and maintaining large-scale, standing participant panels is resource‑intensive. FRAME’s goal is not for every organization to build a permanent real‑world AI evaluation lab but to function as centralized infrastructure, absorbing the burden of recruiting, verifying, and running panels at scale. This centralization allows many institutions to draw on human‑grounded evidence while FRAME panels and protocols bear the operational load.

\textbf{Grounded, transparent scoring constructs}
FRAME's contextual scoring engine organizes AI‑in‑use around clearly defined ideas that matter in a given scenario, population, or setting. Rubrics are based on observational data so interactions can first be grouped into descriptive categories, such as: patterns of use, non‑use, reliance, abandonment, and workarounds. System scores then show how often each pattern appears across different groups, tasks, and contexts. Rubrics and scoring rules are documented in public‑facing materials so others can review, question, adapt, and reuse them against their own norms and risk thresholds.

\textbf{Iterative formalization of metrics}
The precise formulas, data schemas, and standards for FRAME’s six metric families are intentionally kept at a high level in this paper. Because real‑world environments are highly varied, any formalization requires ongoing validation. FRAME members refine developed methods iteratively and, as metrics are validated, publish them to build a transparent, shared catalog that can serve as a common language for the field.

\textbf{Fixed benchmark datasets vs dynamic traces}
Notably, FRAME’s sandbox does not rely on fixed benchmark datasets. Instead, it collects new data directly from the sandbox, then folds observational traces into shared community models. This keeps the evidence base tied to how people actually use AI systems over time. Dynamic traces can complement fixed benchmarks by clarifying how model capabilities materialize—or fail to—in specific contexts.

\textbf{Protecting privacy via contextual mimicry}
Measuring real operational friction often raises concerns about access to proprietary systems or private employee data. FRAME does not tap into organizations’ internal data streams. Instead, it works with sectors to map constraints and workflows, then recreates those environments inside the sandbox using proxy scenarios and specialized panels (for example, credentialed nurses or civil servants). This approach is designed to protect sensitive data while still capturing domain‑specific user entropy and norms, rather than relying solely on general‑purpose consumer use in everyday settings.

\section{About the Intiative}
\label{AboutFRAME}

\subsection{Origins and governance}
FRAME emerged from the AI Metrology Working Group (AI MWG), launched in summer 2024 by Humane Intelligence under the leadership of Dr. Rumman Chowdhury. The AI MWG was a dedicated community focused on strengthening the scientific measurement of AI and expanding evaluation capacity beyond big tech. Over time, this effort evolved into FRAME, whose members span industry, government, civil society, and academia. FRAME is managed by Civitaas Insights and anchored at Virginia State University’s Center for Responsible AI, which serves as host and sponsor rather than a commercial beneficiary. This structure is designed to safeguard independence while providing stable governance, administrative support, and robust conflict‑of‑interest protections.

\subsection{Methodological Foundations} 
FRAME adapts established scientific traditions—program evaluation \cite{scriven_evaluation_1991}, realist evaluation\cite{pawson_introduction_1997,manzano_realist_2025,nielsen_unpacking_2022,louart_realist_nodate}, and implementation science \cite{barry_creating_2025,eccles_welcome_2006}—to the challenges of AI measurement. This orientation allows FRAME to: 
\begin{itemize}
    \item treat AI deployment as an intervention in a complex socio-technical system,
    \item move beyond abstract performance scores to uncover "what works, for whom, in what circumstances”\cite{palm_what_2020,punton_reality_2020,greenhalgh_understanding_2022,harris_forgotten_2025},
    \item leverage rigorous quasi‑experimental designs in real‑world testing \cite{campbell_methods_1991,shadish_experimental_2002}, and
    \item build up the causal evidence decision-makers need to distinguish genuine utility from merely theoretical capability.
\end{itemize}

\subsection{Engaging with FRAME}
Organizations and communities can work with FRAME to access empirical evidence grounded in settings like their own and to help resolve the “decision‑maker’s dilemma.” Through paid tiers,\footnote{FRAME maintains a strict firewall between sponsors and its scientific work. Sponsors can pose questions and fund studies, but all methods, analysis, and publication decisions follow VSU’s policies, not sponsor preferences.} partners support the initiative and receive tailored, decision‑ready insights tied to their specific use cases and domains:

\begin{itemize}
    \item \textbf{Sponsorship \& targeted trials}: Organizations can underwrite sandbox trials to evaluate AI systems in their own use cases (for example, a benefits chatbot or a newsroom tool) under conditions that reflect how people actually adopt, adapt, or abandon systems.
    \item \textbf{Specialized panels}: FRAME collaborates with sponsors to curate specialized participant panels—such as educators or defined consumer segments—organized as short‑term or longitudinal cohorts so results reflect the populations and contexts that matter most.
    \item \textbf{Licensing \& decision‑ready evidence}: Sponsors can license access to FRAME’s community models and detailed metrics to compare their own pilots against broader patterns of risk and value—without sharing or exposing their proprietary data and systems.
\end{itemize}

\section*{Acknowledgments}
This white paper was developed as part of the Forum for Real‑World AI Measurement and Evaluation (FRAME) at Virginia State University’s Center for Responsible AI.

The authors would like to thank members of the AI Metrology Working Group who provided input on FRAME's structure and focus: Morgan Briggs, Peter Douglas, Matt Holmes, Fariza Rashid, Isar Nejadgholi, Afaf Ta\"ik, Carina Westling, and Kyra Wilson. The authors also thank Sundeep Bhandari and Rumman Chowdhury for their feedback and suggestions on early drafts of this document. Thanks to the entire Center for Responsible AI team at VSU: Maurice B. Jones, Sylvia Jones, Dennis Donaldson, PhD. (affiliate researcher), and M. Omar Faison, PhD (affiliate researcher).

\section*{AI Use Disclosure}
Portions of this white paper were developed with the assistance of generative AI tools. These tools were used to help format the LaTeX/Overleaf layout, streamline and clarify draft text, suggest alternative phrasings, assist in generating illustrative figures based on author requirements, and create or refine a subset of the BibTeX entries. All ideas, arguments, structure, and final wording were reviewed, edited, and approved by the authors, who take full responsibility for the content.

\onecolumn
\section*{Glossary of Real-World AI Evaluation Terminology}
\begin{itemize}
    \item Abstract knowledge: Information about what a model can do in principle, usually measured in controlled or lab-like settings.
    \item Accessibility: Whether a system is usable across different abilities, languages, and cultural backgrounds, and whether it avoids creating new barriers for specific groups.
    \item Adaptability: How well a system or workflow can adjust to new users, contexts, tasks, or constraints without breaking or requiring extensive rework.
    \item AI deployment: Phase of a project where a system is put into operation and cutover issues are resolved\cite{ISOIECIEEE24765_2017}.
    \item AI model capabilities: The tasks, functions, and behaviors an AI model can reliably perform, given specified inputs and conditions.
    \item AI stack: A layered set of technologies, tools, frameworks, and infrastructure for building, deploying, and operating AI systems and applications.
    \item AI system: A machine-based system that, for explicit or implicit objectives, infers, from the input it receives, how to generate outputs such as predictions, content, recommendations, or decisions that can influence physical or virtual environments. Different AI systems vary in their levels of autonomy and adaptiveness after deployment\cite{OECD2024AISystemDefinition}.
    \item Assessment: Action of applying specific documented criteria to a specific software module, package or product for the purpose of determining acceptance or release of the software module, package or product\cite{ISOIECIEEE24765_2017}.
    \item Availability: Ensuring timely and reliable access to and use of information\cite{nist_sp800-37r2}.
    \item Benchmarking: A systematic method by which organizations can measure themselves against the best industry practices.
    \item Community models: Digital snapshots of how AI is actually used in a given setting —capturing user behavior, tasks, failure modes, and higher‑order effects—so organizations can compare systems, test policies, and tailor local evaluations without building full pipelines.
    \item Construct: An abstract, latent concept or theoretical variable, not directly observable, that is defined for scientific purposes and measured indirectly through multiple observable indicators or items.
    \item Construct operationalization: Linking a systematized concept to appropriate indicators and scores in a coherent measurement framework\cite{Adcock_Collier_2001}.
    \item Construct systematization: The process of clarifying and organizing a concept by specifying its meaning, dimensions, and relationships to other concepts.\cite{Adcock_Collier_2001}.
    \item Context: The parameters in which interrelated factors, purposes, and circumstances may shape individual and collective perceptions, interpretations, and expectations about the functionality and impacts of AI technology, and resulting actions\cite{nist_aria_companion_2020}.
    \item Contextual knowledge: Detailed insight tied to a specific setting, including the local conditions, people, practices, and history that give information its meaning and practical relevance in that context.
    \item Decision-maker’s dilemma: The predicament faced by stakeholders outside the technical stack—such as CIOs, agency heads, editors, and compliance officers—who must decide whether and how to deploy AI systems without solid evidence about how those systems actually behave in the environments they oversee.
    \item Decision-ready evidence: Systematic indicators derived from observing AI-in-use that help stakeholders outside the stack judge utility, risk, and operational value in their own context, so they can act on procurement, deployment, governance, and evaluation.
    \item Descriptivist methods: An evidence‑based approach to language that describes how people actually use it, rather than telling them how they should use it.
    \item Evaluation: (1) Systematic determination of the extent to which an entity meets its specified criteria; (2) Action that assesses the value of something\cite{ISOIECIEEE24765_2017}.
    \item Evaluation scenario: A high-level description of a specific situation or sequence of conditions under which a system is to be tested, including the initial state, inputs, triggering events, and expected behavior or outcomes.
    \item Generative AI: A category of AI that can create new content such as text, images, videos and music\cite{OECD2023GenerativeAI}.
    \item Ground truth: Value of the target variable for a particular item of labelled input data\cite{iso_iec_22989_2022}.
    \item Higher‑order effects: Long-term and broad-scale outcomes and consequences that may result from AI use in the real world.
    \item In-silico testing: Testing or experimentation carried out entirely on a computer, using computational models and simulations.
    \item Interoperability: Degree to which two or more systems, products or components can exchange information and use the information that has been exchanged\cite{ISOIECIEEE24765_2017}.
    \item Measurement: (1) Quantitative measurement is the act or process of assigning a number or category to an entity to describe an attribute of that entity\cite{ISOIECIEEE24765_2017}. (2) Qualitative measurement is based on descriptive data such as through observations, interviews, focus groups, or open-ended text fields in surveys.
    \item Model stochasticity: The inherent randomness in generative model outputs that makes the same or similar prompt produce different answers on different runs..
    \item Predictive AI: AI systems whose primary function is to infer from input data and produce predictions or forecasts about future states, behaviors, or outcomes, typically to inform decisions or actions \cite{OECD2024AISystemDefinition}.
    \item Privacy: Freedom from intrusion into the private life or affairs of an individual\cite{iso_iec_ts_5723_2022}. 
    \item Proxy scenario: A test scenario that does not directly reproduce the real-world context of interest but is designed to stand in for it, using more tractable or safer conditions while preserving key features believed to be relevant for evaluating system behavior or risk.
    \item Quality: The totality of features and characteristics of a product or service that bear on its ability to satisfy stated or implied needs \cite{oecd_glossary_2008}
    \item Real‑world AI evaluation: The process of measuring what actually happens when people use AI systems in everyday workflows and contexts, at a scale that allows patterns to emerge across different users and settings.
    \item Reality Gap: The difference between a model’s performance in optimized test conditions and the outcomes it produces with real people in real contexts.
    \item Resilience: The ability of an information system to continue to: (i) operate under adverse conditions or stress, even if in a degraded or debilitated state, while maintaining essential operational capabilities; and (ii) recover to an effective operational posture in a time frame consistent with mission needs\cite{nist_sp800-39}.
    \item Robustness: Ability of a system to maintain its level of performance under a variety of circumstances\cite{iso_iec_ts_5723_2022}.
    \item Situated observation: The practice of observing activity directly in its natural setting and attending to how behavior unfolds over time within specific physical, social, and cultural environments.
    \item Systematic knowledge: Organized, cumulative knowledge that a community maintains and updates so that its concepts, measures, and explanations can be applied across many contexts, not just a single local case.
    \item Testing, Evaluation, Validation, and Verification (TEVV): A framework for assessing and incorporating methods and metrics to determine that a technology or system satisfactorily meets its design specifications and requirements, and that it is sufficient for its intended use \cite{NSCAI2021FinalReport}.
    \item User entropy: The inherent heterogeneity in how people use AI in context, including how they express needs, interpret and adapt outputs, and embed AI into their own goals and constraints.
\end{itemize}

\printbibliography
\end{document}